\documentclass[conference]{IEEEtran}
\IEEEoverridecommandlockouts
\usepackage{cite}
\usepackage{amsmath,amssymb,amsfonts}
\usepackage{algorithmic}
\usepackage{graphicx}
\usepackage{subcaption}
\usepackage{caption}
\usepackage{float}
\usepackage{textcomp}
\usepackage{makecell}
\usepackage{booktabs}
\usepackage{xcolor}
\usepackage{url}
\def\BibTeX{{\rm B\kern-.05em{\sc i\kern-.025em b}\kern-.08em
    T\kern-.1667em\lower.7ex\hbox{E}\kern-.125emX}}
\begin{document}

\title{ICME 2025 Generalizable HDR and SDR Video Quality Measurement Grand Challenge}




\author{
Yixu Chen$^*{\dagger}$, \and
Bowen Chen$^*{\dagger}$, \and
Hai Wei$^{\dagger}$, \and
Alan C. Bovik$^{\dagger}$,
\and
Baojun Li$^{\ddagger}$, \and
Wei Sun$^{\S}$, \and
Linhan Cao$^{\S}$, \and
Kang Fu$^{\S}$, \and
Dandan Zhu$^{\S}$, \and
Jun Jia$^{\S}$, \and
Menghan Hu$^{\S}$, \and
Xiongkuo Min$^{\S}$, \and
Guangtao Zhai$^{\S}$, \and
Dounia Hammou$^{\parallel}$, \and
Fei Yin$^{\parallel}$, \and
Rafał K. Mantiuk$^{\parallel}$, \and
Amritha Premkumar$^{\P}$, \and
Prajit T Rajendran$^{\P}$, \and
Vignesh V Menon$^{\P}$, \and
}

\maketitle

\let\thefootnote\relax\footnotetext{The challenge is sponsored by Amazon Prime Video. The organizers are noted by $^{\dagger}$. These authors* contributed equally. The other authors who win the challenge and their teams are indicated by the following team names: $^{\ddagger}$ SLCV, $^{\S}$ ECNU-SJTU VQA, $^{\parallel}$ Cambridge VDP, $^{\P}$ PhoenixVideo. Website: \url{https://sites.google.com/view/icme25-vqm-gc/}. Code: \url{https://github.com/Easoncyx/vqm_challenge}.}

\begin{abstract}

This paper reports IEEE International Conference on Multimedia \& Expo (ICME) 2025 Grand Challenge on Generalizable HDR and SDR Video Quality Measurement. With the rapid development of video technology, especially High Dynamic Range (HDR) and Standard Dynamic Range (SDR) contents, the need for robust and generalizable Video Quality Assessment (VQA) methods has become increasingly demanded. Existing VQA models often struggle to deliver consistent performance across varying dynamic ranges, distortion types, and diverse content. This challenge was established to benchmark and promote VQA approaches capable of jointly handling HDR and SDR content. In the final evaluation phase, five teams submitted seven models along with technical reports to the Full Reference (FR) and No Reference (NR) tracks. Among them, four methods outperformed VMAF baseline, while the top-performing model achieved state-of-the-art performance, setting a new benchmark for generalizable video quality assessment. 

\end{abstract}


\section{Introduction}

Video Quality Assessment (VQA) focuses on understanding how humans perceive the quality of videos affected by distortions introduced during acquisition, compression, transmission, or enhancement. VQA plays a key role in various video processing applications such as video coding, restoration, and streaming. Despite significant progress, accurately predicting human-perceived video quality remains difficult due to the various distortions, content characteristics, and temporal dynamics.

The growing development of High Dynamic Range (HDR) video, alongside traditional Standard Dynamic Range (SDR) content, introduces further challenges for VQA. HDR and SDR differ in key visual properties such as peak luminance, and color gamut, which influence how distortions are perceived. Many current VQA models are designed specifically for either HDR or SDR, limiting their effectiveness when applied across both formats. These limitations are even more noticeable when evaluating a variety of distortions or diverse content types.

To address these issues, the ICME 2025 Grand Challenge on Generalizable HDR and SDR Video Quality Measurement was introduced. The goal is to benchmark VQA methods that can perform consistently across both HDR and SDR content. The challenge includes two tracks: Full-Reference, where reference videos are available, and No-Reference, where the model must predict quality without access to the original video. These two tracks simulate real-world conditions ranging from controlled media production environments to open, user-generated content scenarios. An HDR\&SDR VQA database is used for this challenge, covering a wide range of content types and distortion conditions in both HDR and SDR formats. Subjective quality scores were collected for evaluation as the ground truth. 

The models proposed by different participating teams demonstrate a range of strategies, including large multimodal foundational models tuned for VQA tasks, human vision system modeling with transformer, traditional interpretable coding features and machine learning pipelines. Several teams addressed the limited availability of HDR training data through pretraining on SDR datasets and domain adaptation techniques. Others introduced customized feature extraction, temporal modeling, or data augmentation strategies to enhance generalization across content types and distortion levels. This diversity reflects ongoing trends in video quality assessment: combining perceptual modeling, data-driven learning, and practical considerations. 

In total, five teams submitted their final solutions to the challenge, spanning FR and NR settings. Four out of seven models outperformed VMAF baseline, and the top-performing method achieved state-of-the-art results under the official evaluation protocol. This paper presents the design of the challenge, details of the dataset, evaluation methodology, and an overview of the submitted methods and their performance.

\section{Challenge Design}

We organize the ICME 2025 Grand Challenge on Generalizable HDR and SDR Video Quality Assessment to facilitate the development and benchmarking of perceptual VQA models that perform consistently across HDR and SDR formats. The challenge is structured to address three core objectives: (1) encouraging novel VQA models capable of handling both HDR and SDR content under a variety of distortion conditions; (2) providing a standardized benchmark for both FR and NR methods; and (3) releasing an open-sourced large-scale HDR\&SDR video quality dataset with high-quality subjective labels to support future research.

\textbf{Task:} Participants are tasked with designing models that predict quality scores in the range [0, 10], aiming for strong alignment with human perceptual judgments. Two tracks are provided:
\begin{itemize}
    \item Full-Reference Track: Models are given access to both the original and distorted versions of each video.
    \item No-Reference Track: Models must operate using only the distorted input.
\end{itemize}

Submissions must support both SDR and HDR10 video formats.

\textbf{Dataset:} The challenge uses a subset of the HDRSDR-VQA dataset~\cite{chen2025hdrsdrvqasubjectivevideoquality} developed by the LIVE Lab at the University of Texas at Austin, in collaboration with Amazon Prime Video. This subset consists of 31 open-source source videos selected from the 8K HDR AVT-VQDB-UHD-2-HDR dataset~\cite{AVT-VQDB-UHD-2-HDR}. Each sequence was encoded in both HDR10 (BT.2020, PQ, yuv420p10le) and SDR (BT.709, yuv420p) formats, with a duration of approximately 7 seconds per clip.

HDR-to-SDR conversions were performed using NBCUniversal LUTs to ensure consistent tone-mapping across sequences. Each source video was further encoded at nine bitrate and resolution levels, resulting in 18 distorted versions per content (9 HDR, 9 SDR). Encoding was carried out using FFmpeg with the libx265 encoder in constant bitrate mode.

All videos are annotated using subjective scores collected through a pairwise comparison study conducted on six HDR10-capable consumer displays. The study followed ITU-R BT.500.13~\cite{bt500} and used the Active Sampling for Pairwise Comparisons (ASAP) algorithm ~\cite{asap}. The collected data were scaled into continuous Just-Objectionable-Difference (JOD) scores with the \textit{pwcmp} algorithm~\cite{pwcmp}.

For the challenge, 20 open-source videos (360 sequences: 180 HDR and 180 SDR) are released for training. The remaining 11 contents (198 sequences: 99 HDR and 99 SDR) are reserved for the test set. In the FR track, the 3840×2160 versions encoded at 50 Mbps are used as the reference videos.

\textbf{Evaluation:} Submissions are evaluated by comparing the predicted quality scores against the ground-truth JOD values derived from subjective experiments. The primary evaluation metric is Spearman’s Rank Order Correlation Coefficient (SROCC), which assesses the monotonic relationship between predicted and subjective scores.

In addition to SROCC, the following metrics are also reported:
\begin{itemize}
    \item Pearson’s Linear Correlation Coefficient (PLCC), computed after a third-order polynomial regression to account for possible non-linear relationships.
    \item Root Mean Square Error (RMSE), also calculated post-regression, to reflect absolute prediction error.
    \item Kendall’s Rank Correlation Coefficient (KROCC), which provides an additional non-parametric measure of ranking consistency.
    \item Inference time, the average inference time  across the test set with a single Nvidia L40S GPU.
\end{itemize}

Before calculating the PLCC and RMSE, a four-parameter non-linear mapping \cite{non_linear} is used to map the model's prediction to MOS on private test set to account for scale difference.
$$ f(o) = \frac{\beta_1-\beta_2}{1+e^{-\frac{o-\beta_3}{|\beta_4|}}} +\beta_2 $$ 
where o is the objective model prediction.

Runtime is measured on an AWS EC2 G6e.8xlarge instance. It is not used for ranking, but is included to contextualize model complexity.

\textbf{Challenge Phases and Submission:} During the development phase, participants are provided with the training data along with the corresponding ground-truth JOD scores to facilitate model training and validation. In the final testing phase, models are evaluated on the unseen test set. Participants are required to submit their final models packaged in Docker containers to ensure reproducibility and standardized evaluation.


\section{Challenge Results}

\begin{table*}[ht]
\centering
\caption{\centering Challenge results across all, SDR, and HDR10 videos. Metrics shown: SROCC / PLCC / KROCC / RMSE. \textit{Inf Time} denotes the average inference time per video across the entire test set on Nvidia L40S GPU unless noted otherwise.}
\begin{tabular}{llcccc}
\toprule
\textbf{Track} & \textbf{Team/Model Name} & \textbf{All} & \textbf{SDR} & \textbf{HDR10} & \textbf{Inf Time (s)} \\
\midrule
NR & SLCV~\cite{slcv}               & \makecell{0.945 / 0.943 / 0.802 / 0.465} & \makecell{0.933 / 0.943 / 0.787 / 0.450} & \makecell{0.953 / 0.944 / 0.823 / 0.476} & 42.27\\
FR & ECNU-SJTU-FR~\cite{ecnu}            & \makecell{0.932 / 0.941 / 0.778 / 0.482} & \makecell{0.924 / 0.944 / 0.767 / 0.454} & \makecell{0.938 / 0.939 / 0.789 / 0.507} & 17.34 \\
FR & cvvdpMlTransformer~\cite{colorvideo-ml} & \makecell{0.917 / 0.910 / 0.757 / 0.580} & \makecell{0.908 / 0.914 / 0.751 / 0.550} & \makecell{0.927 / 0.906 / 0.769 / 0.609} & 76.96\\
FR & cvvdpMlSaliency~\cite{colorvideo-ml}    & \makecell{0.910 / 0.908 / 0.743 / 0.588} & \makecell{0.923 / 0.923 / 0.769 / 0.521} & \makecell{0.897 / 0.894 / 0.726 / 0.647} & 76.87 \\
NR & ECNU-SJTU-NR~\cite{ecnu}            & \makecell{0.880 / 0.897 / 0.697 / 0.632} & \makecell{0.843 / 0.882 / 0.657 / 0.650} & \makecell{0.922 / 0.914 / 0.757 / 0.604} & 17.84 \\
FR(RR) & PhoenixVideo (CPU)~\cite{phenixVideo}       & \makecell{0.832 / 0.787 / 0.632 / 0.862} & \makecell{0.871 / 0.846 / 0.677 / 0.719} & \makecell{0.800 / 0.760 / 0.595 / 0.936} & 10.01 \\
NR & BVI-VQA     & \makecell{0.603 / 0.579 / 0.422 / 1.139} & \makecell{0.545 / 0.557 / 0.384 / 1.121} & \makecell{0.659 / 0.613 / 0.467 / 1.138} & 22.74\\
\midrule
\multicolumn{5}{l}{\textbf{Baseline Methods on CPU}} \\
\midrule
NR & P1204.3 (1 thread)~\cite{rao2020p1204}     & \makecell{0.925 / 0.935 / 0.762 / 0.496} & \makecell{0.928 / 0.934 / 0.768 / 0.484} & \makecell{0.927 / 0.945 / 0.775 / 0.470} & 121.24\\
NR & EQM decoder (12 threads)~\cite{eqm}	& \makecell{0.925 / 0.925 / 0.762 / 0.531} & \makecell{0.919	/ 0.917	/ 0.751 / 0.538} & \makecell{0.939 / 0.950 / 0.795 / 0.448} & 16.65\\
FR & VMAF (12 threads)~\cite{vmaf}         & \makecell{0.905 / 0.901 / 0.725 / 0.597} & \makecell{0.922 / 0.915 / 0.755 / 0.541} & \makecell{0.896 / 0.909 / 0.720 / 0.594} & 83.11\\
FR & PSNR-Y       & \makecell{0.674 / 0.677 / 0.480 / 1.015} & \makecell{0.727 / 0.737 / 0.542 / 0.901} & \makecell{0.707 / 0.750 / 0.522 / 0.937} & - \\
\bottomrule
\end{tabular}
\label{tab:challenge_results_with_track}
\end{table*}

\begin{figure*}[ht]
    \centering
    \begin{subfigure}[b]{0.23\textwidth}
        \includegraphics[width=\textwidth]{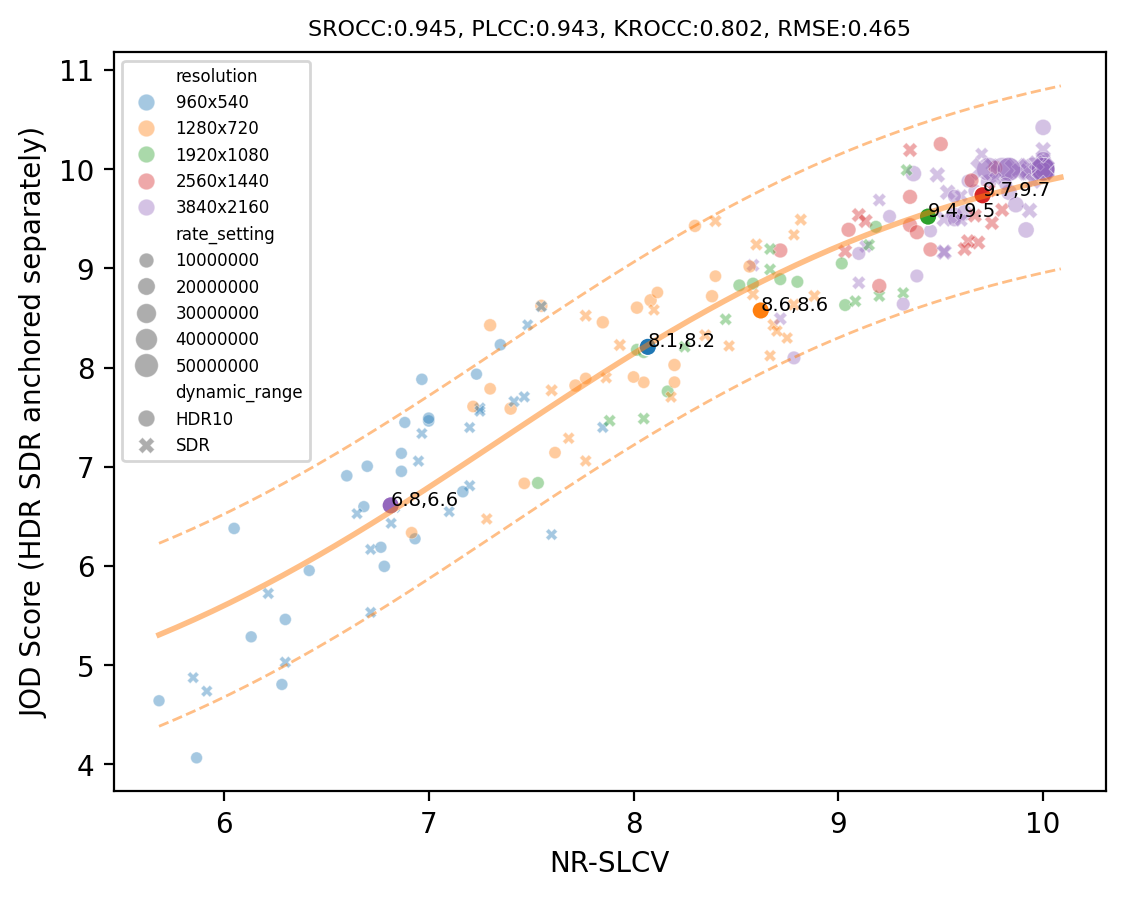}
        \caption{NR-SLCV-SJTU~\cite{slcv}}
    \end{subfigure}
    \begin{subfigure}[b]{0.23\textwidth}
        \includegraphics[width=\textwidth]{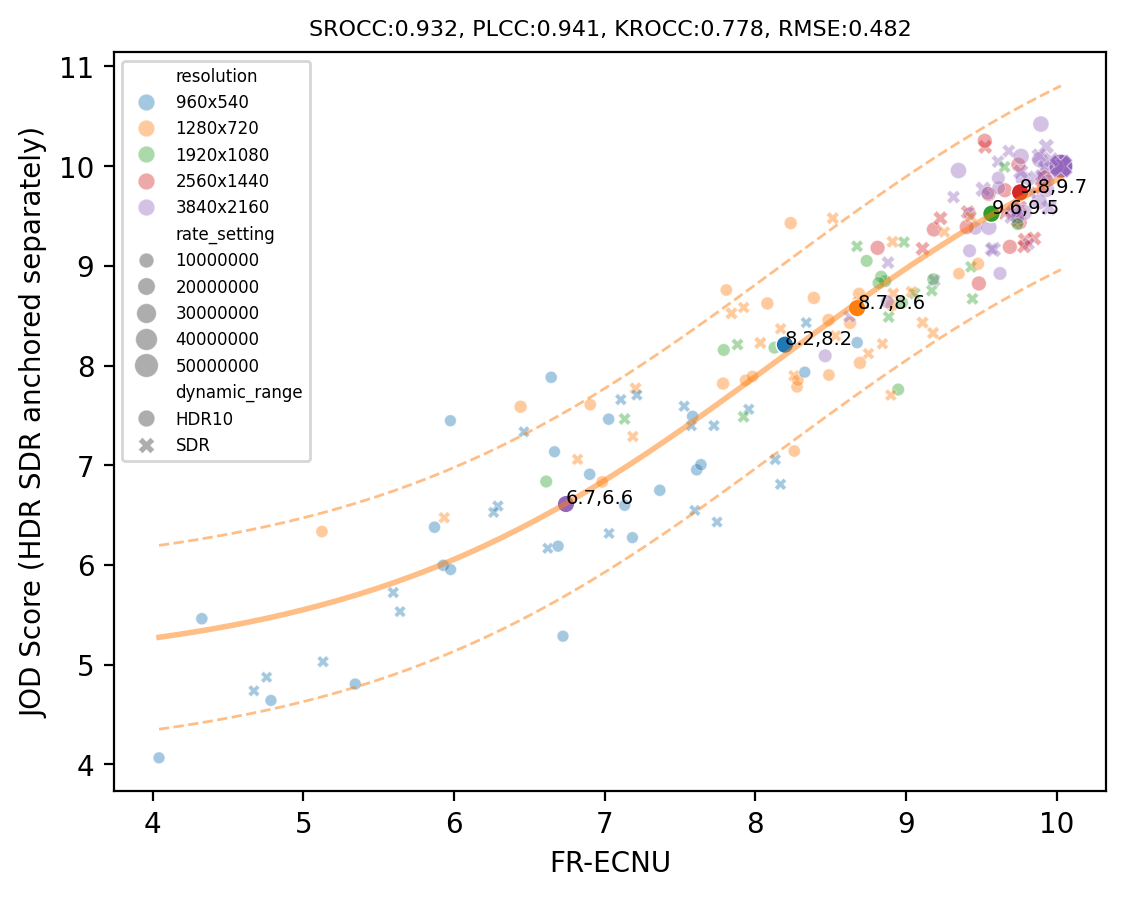}
        \caption{FR-ECNU~\cite{ecnu}}
    \end{subfigure}
    \begin{subfigure}[b]{0.23\textwidth}
        \includegraphics[width=\textwidth]{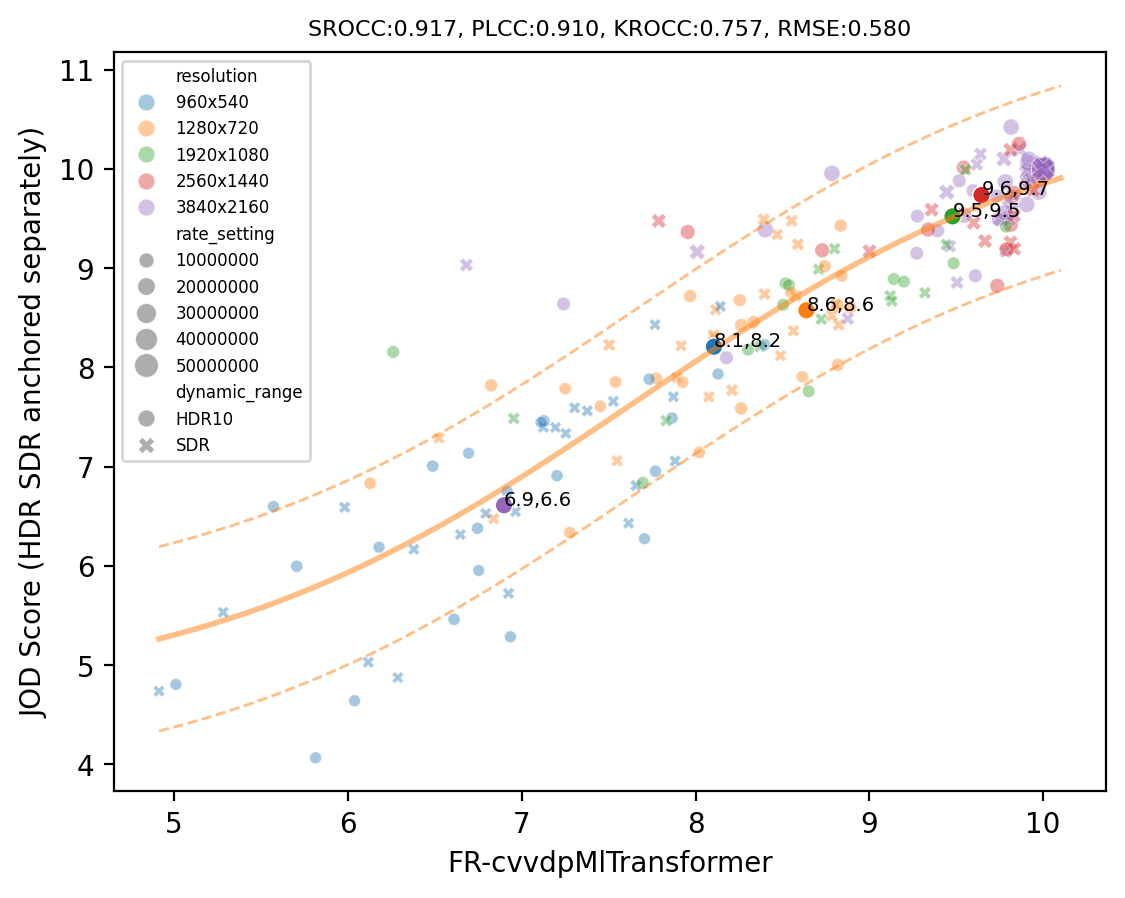}
        \caption{FR-cvvdpMlTransformer~\cite{colorvideo-ml} }
    \end{subfigure}
    \begin{subfigure}[b]{0.23\textwidth}
        \includegraphics[width=\textwidth]{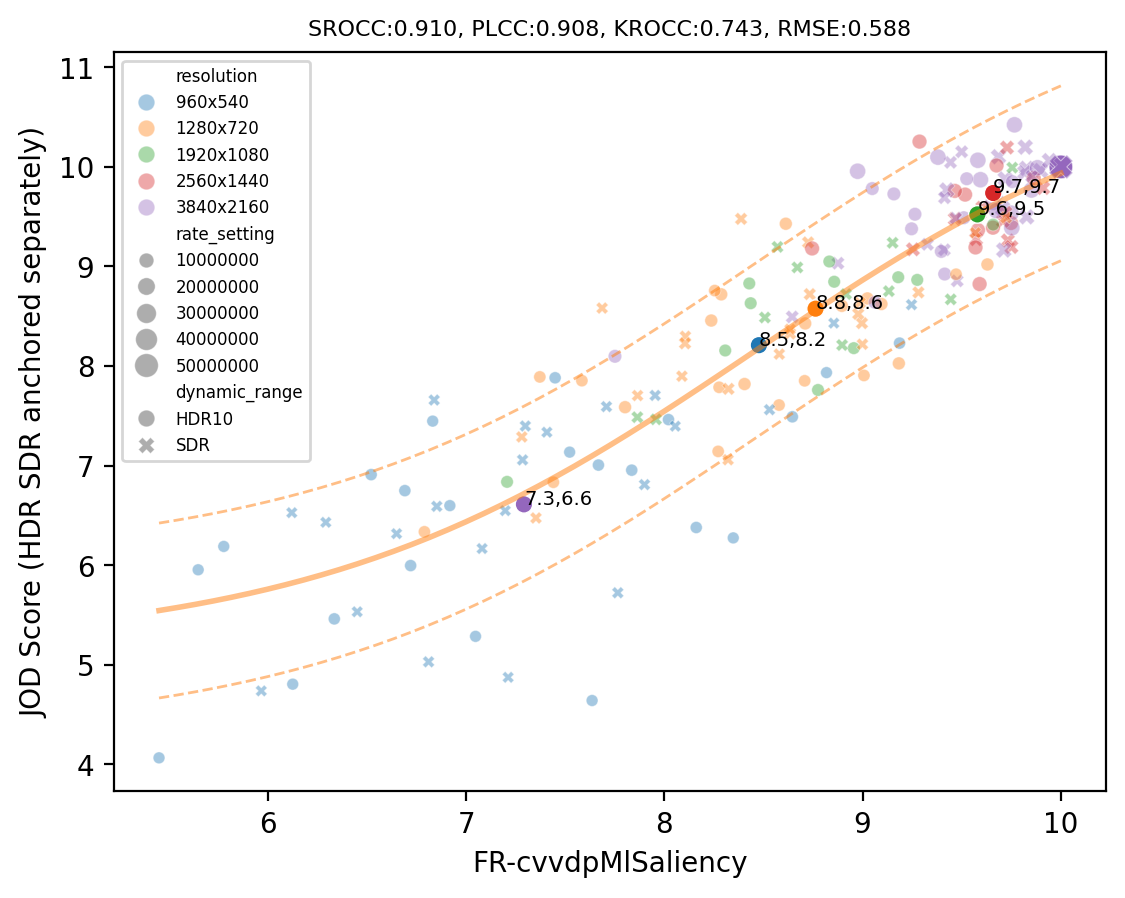}
        \caption{FR-cvvdpMlSaliency~\cite{colorvideo-ml} }
    \end{subfigure}
    
    \vspace{0.3cm}
    
    \begin{subfigure}[b]{0.23\textwidth}
        \includegraphics[width=\textwidth]{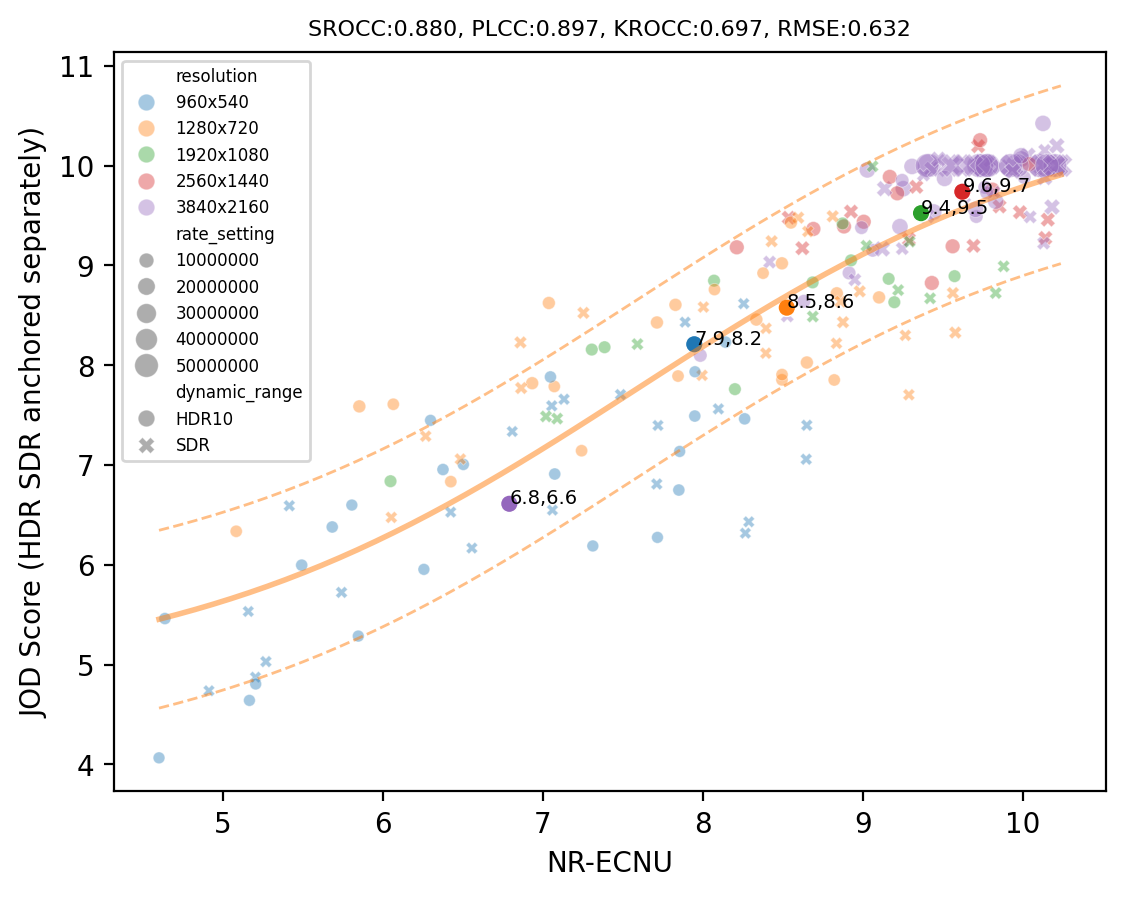}
        \caption{NR-ECNU-SJTU~\cite{ecnu}}
    \end{subfigure}
    \begin{subfigure}[b]{0.23\textwidth}
        \includegraphics[width=\textwidth]{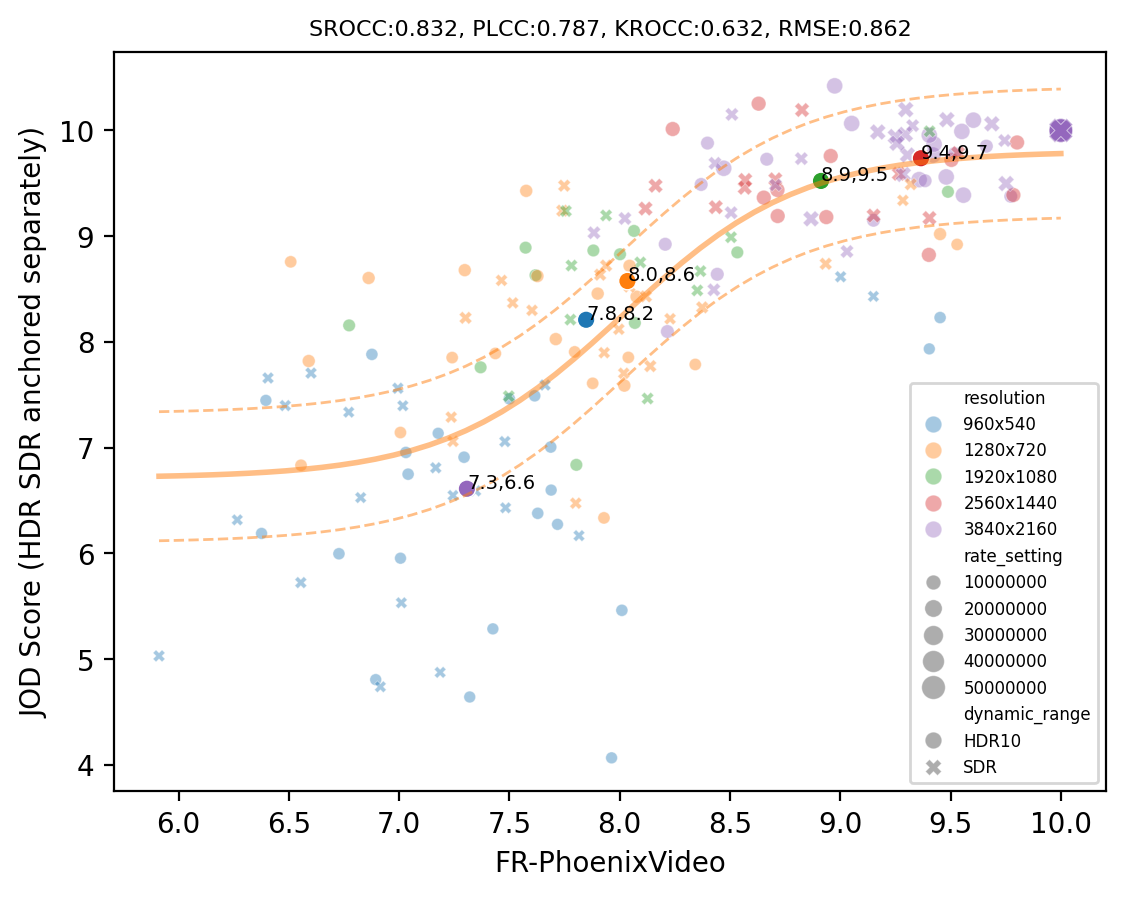}
        \caption{FR-PhoenixVideo~\cite{phenixVideo} }
    \end{subfigure}
    \begin{subfigure}[b]{0.23\textwidth}
        \includegraphics[width=\textwidth]{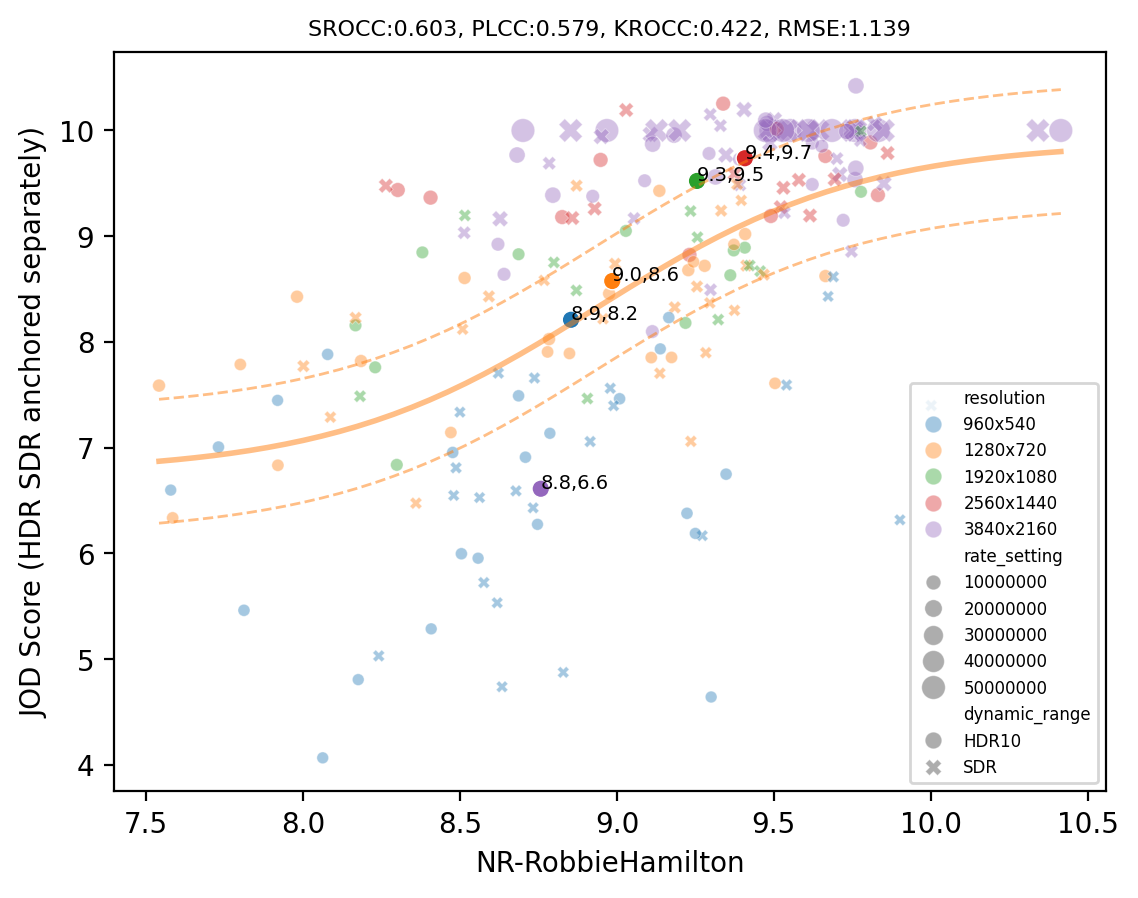}
        \caption{NR-BVI-VQA}
    \end{subfigure}

    \vspace{0.3cm}
        
    \begin{subfigure}[b]{0.23\textwidth}
        \includegraphics[width=\textwidth]{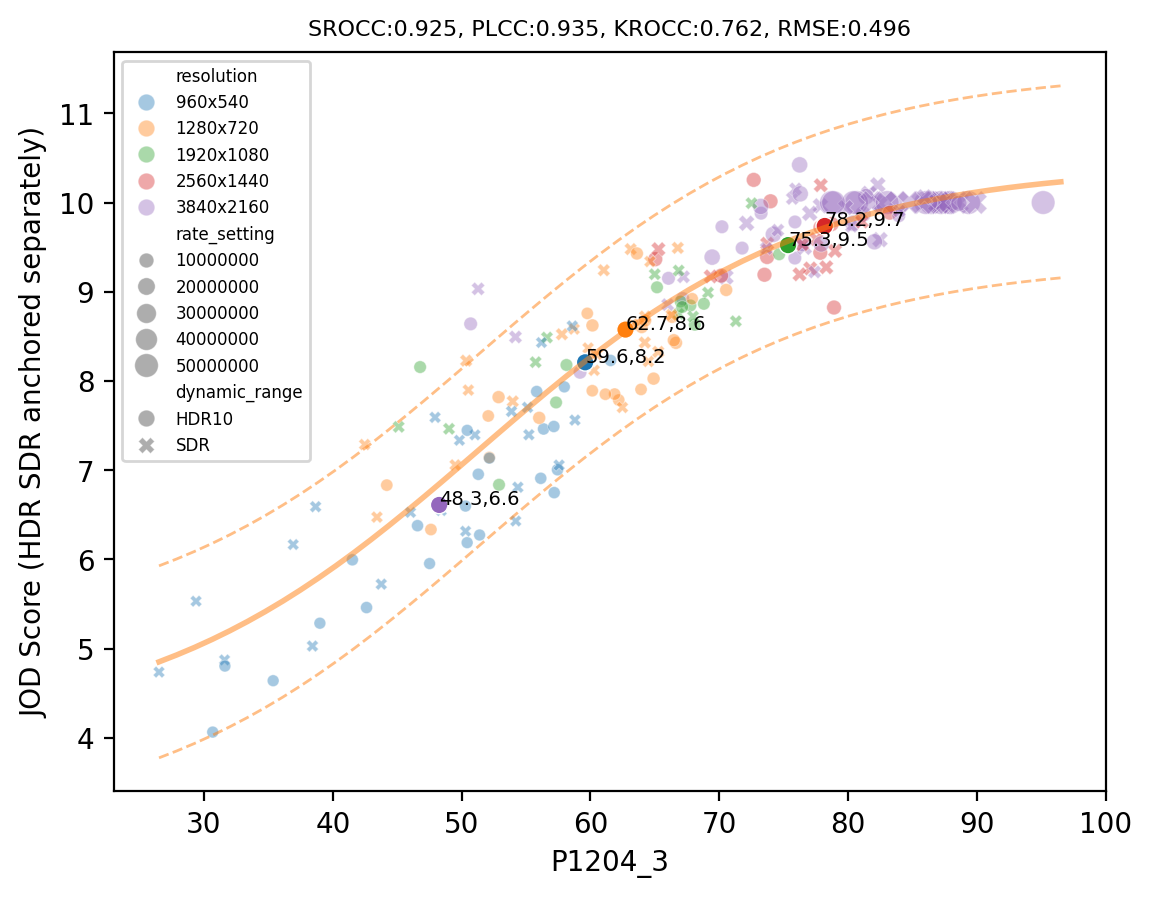}
        \caption{P1204.3 (Baseline)~\cite{rao2020p1204}}
    \end{subfigure}
    \begin{subfigure}[b]{0.23\textwidth}
        \includegraphics[width=\textwidth]{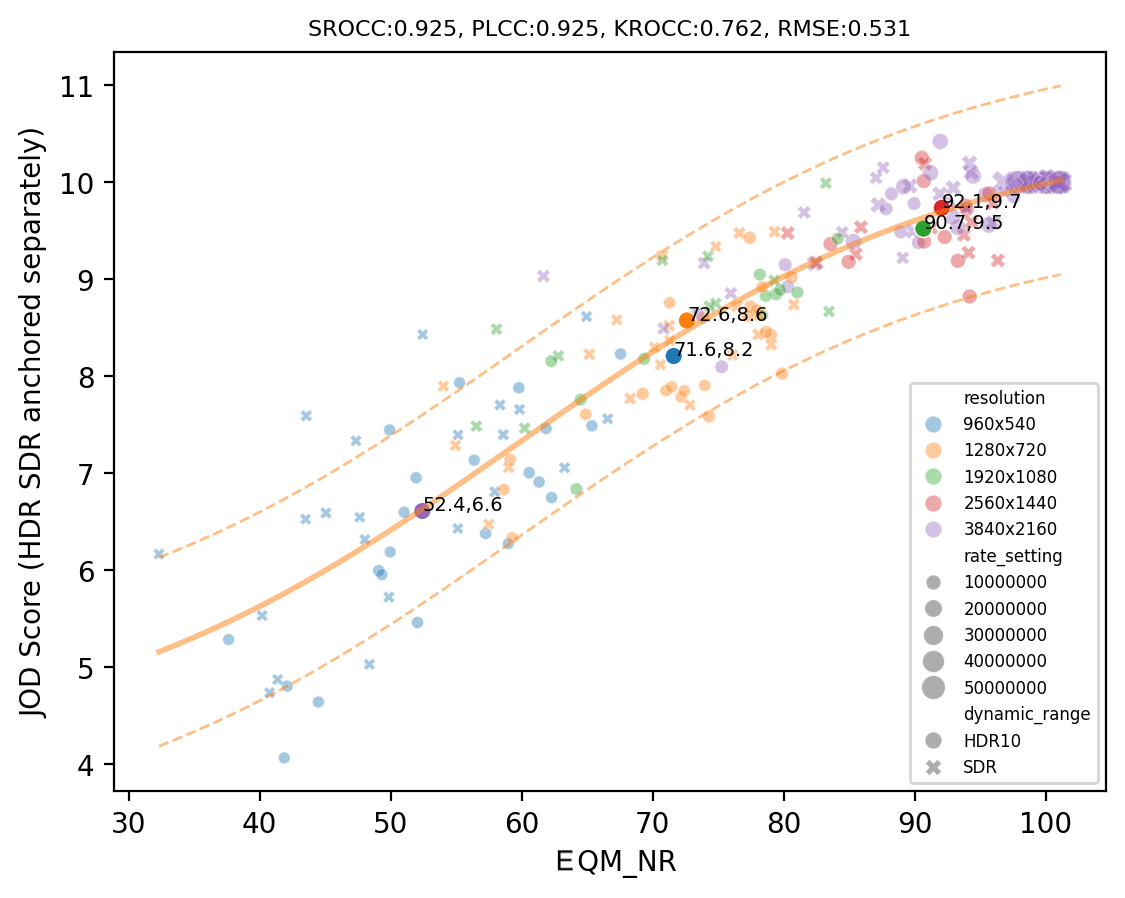}
        \caption{EQM (Baseline)~\cite{eqm}}
    \end{subfigure}
    \begin{subfigure}[b]{0.23\textwidth}
        \includegraphics[width=\textwidth]{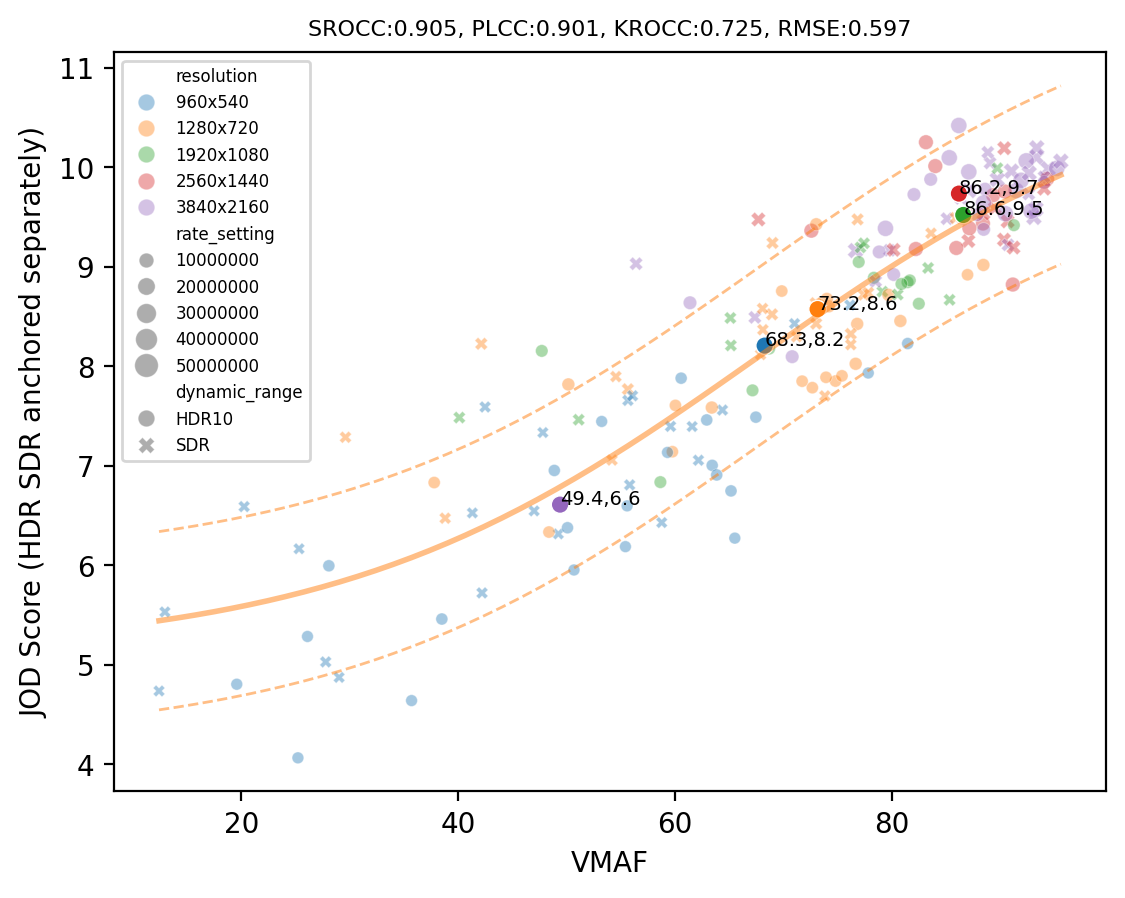}
        \caption{VMAF (Baseline)~\cite{vmaf}}
    \end{subfigure}
    \begin{subfigure}[b]{0.23\textwidth}
        \includegraphics[width=\textwidth]{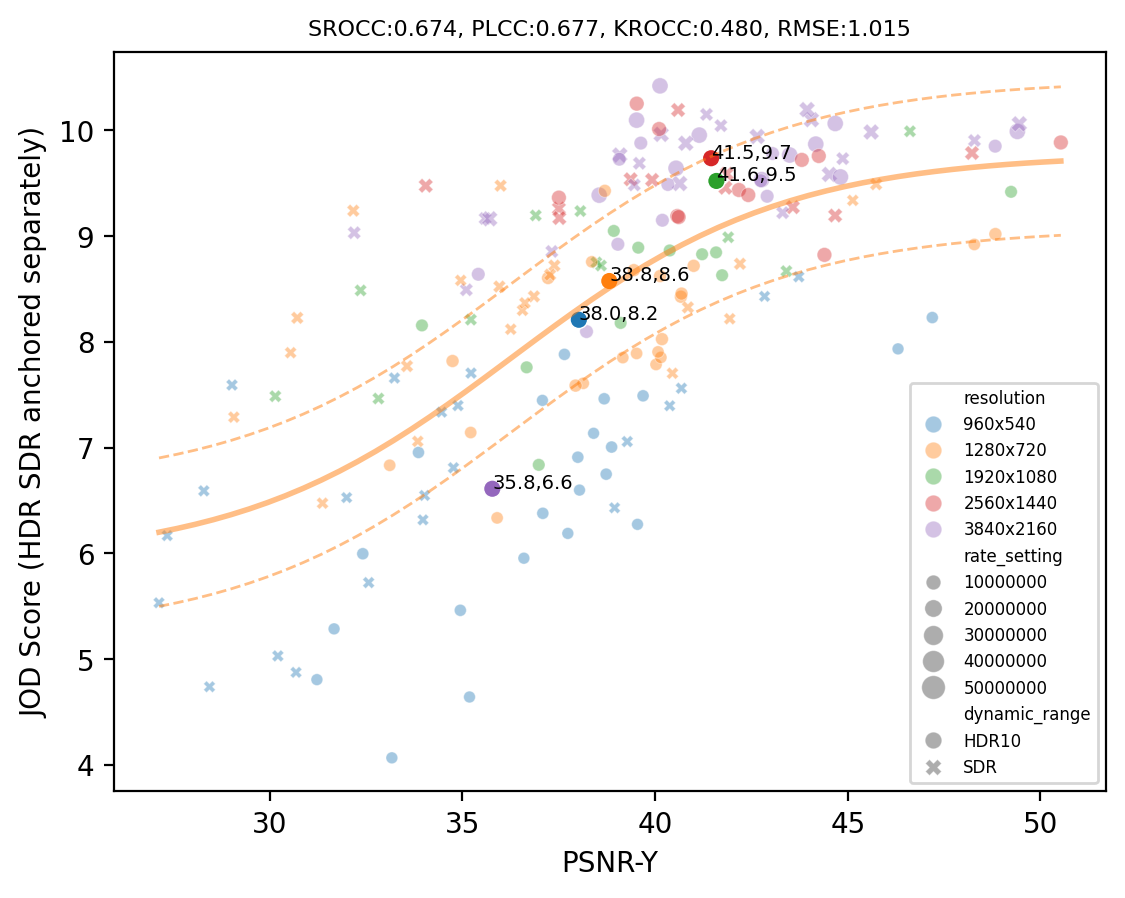}
        \caption{PSNR (Baseline)}
    \end{subfigure}

    \caption{Scatter plots with non-linear mapping between prediction and JOD for all models on the test set. }
    \label{fig:scatter_all_models}
\end{figure*}

Three baseline methods were included to provide reference points for model performance: ITU-T P.1204.3~\cite{rao2020p1204}, EQM~\cite{eqm}, VMAF~\cite{vmaf}, and PSNR-Y. P.1204.3 is a standardized no-reference model developed to assess compressed video quality in streaming scenarios. EQM introduces a lightweight no-reference video quality metric that leverages quantization and motion features extracted from the encoding or decoding process, the speed is measured on the decoder version. VMAF is a widely adopted full-reference metric that combines multiple quality indicators using support vector regression. PSNR-Y is a simple pixel-based full-reference metric that measures the signal fidelity on the luma channel.

Table~\ref{tab:challenge_results_with_track} summarizes the performance of all participating models across the full test set as well as SDR and HDR10 subsets. The table includes standard quality assessment metrics—SROCC, PLCC, KROCC, and RMSE with the average per-video inference time, providing a comprehensive view of both prediction accuracy and computational efficiency.

The top-performing no-reference model, SLCV~\cite{slcv}, achieved the highest overall performance, with an SROCC of 0.945 and PLCC of 0.943, outperforming both traditional baselines and several full-reference models. Among full-reference submissions, ECNU-FR~\cite{ecnu} and cvvdpMlTransformer~\cite{colorvideo-ml} showed strong results, with consistent performance across both SDR and HDR content. Four out of seven participating methods outperformed the official baseline VMAF. All evaluations ran on an AWS EC2 G6e instance. All the baseline models and PhoenixVideo~\cite{phenixVideo} were tested on CPU, while all other submitted models utilized GPU. Fig.~\ref{fig:scatter_all_models} shows the relationship between model-predicted scores and subjective JOD values on the full test set.

\section{Participating Methods}

In this section, we provide detailed descriptions of the top-performing teams in both the FR and NR tracks. We also include brief summaries of the remaining submissions that exceed the baseline to highlight the diversity of approaches explored in the challenge.

\subsection{MLLM-Based NR VQA - Winning Team for NR Track}

\begin{figure*}[!t]
    \centering
    \includegraphics[width=0.8\linewidth]{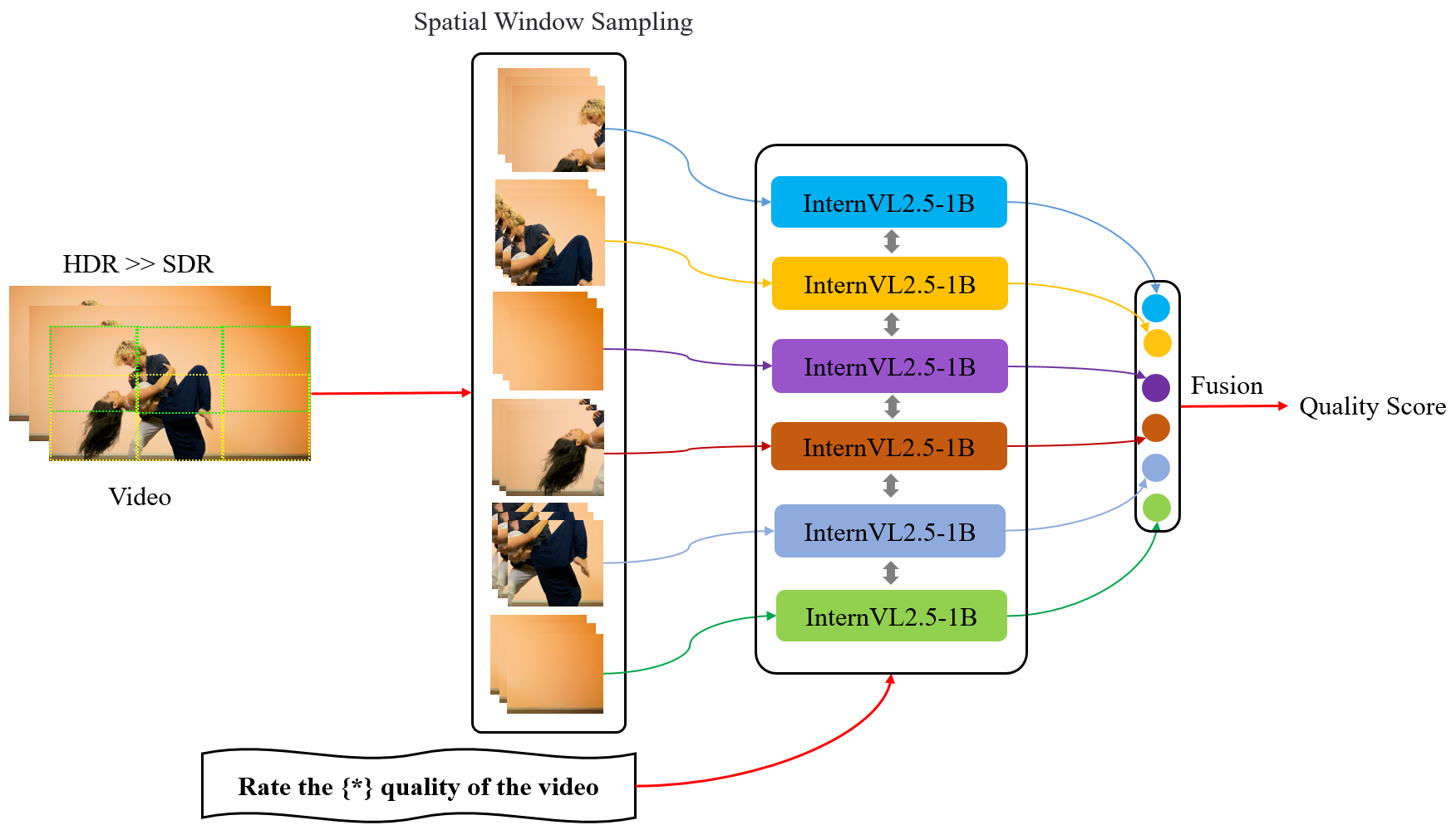}
    \caption{Proposed NR-VQA pipeline based on InternVL 2.5. Tone-mapped HDR/SDR inputs are spatially cropped and passed through a fine-tuned MLLM. Final prediction is the average of regional scores.}
    \label{fig:mlvm-nr}
\end{figure*}

The winning solution~\cite{slcv} for the no-reference (NR) track introduces a novel approach to video quality assessment using a multimodal large language model (MLLM) shown in Fig.~\ref{fig:mlvm-nr}. This method is built upon InternVL 2.5~\cite{internvl}, a vision-language model originally developed for general visual question-answering tasks. The InternVL 2.5 MLLM is adopted as the core architecture for the video quality evaluation pipeline, incorporating FFmpeg-driven HDR-to-SDR preprocessing. This design enables the model to process both HDR and SDR videos within a unified framework, ensuring consistent evaluation of perceptual quality variations while adhering to standard display constraints.

During the preprocessing phase, HDR10 content was converted to SDR equivalents through Mobius tone mapping implemented in FFmpeg (transfer characteristic parameter = 0.3). To maintain perceptual brightness consistency during dynamic range compression, the input signal was linearized under the assumption of a reference display peak luminance of 1000 nits. While this inevitably incurs some information loss inherent to HDR-to-SDR conversion, the standardized processing pipeline ensures consistent input distribution across both HDR and SDR content, thereby facilitating more robust model learning through dynamic range-agnostic feature representation.

The training dataset provided for this NR track contains only 360 videos, which presents significant challenges due to its limited scale. To address this constraint in the spatial domain, a data augmentation strategy called Spatial Window Sampling is introduced. This strategy applies a sliding window approach that crops video frames to a window size equal to two-thirds of the video’s shortest side. This method effectively increases the number of training samples and enhances the model’s ability to learn spatial features at a finer level. The InternVL 2.5 model is fine-tuned using the Low-Rank Adaptation (LoRA)~\cite{lora} technique. 

All experiments were conducted on a computing node equipped with eight NVIDIA A100 GPUs (80 GB memory each) and 1 TB of system RAM. The InternVL2.5-1B model was fine-tuned using LoRA with hyperparameters set to rank=16, alpha=48, and a learning rate of 2e-4. Training was distributed across eight NVIDIA A100-80G GPUs with a batch size of 24 for four epochs to prevent overfitting. During inference, the same sliding window sampling strategy used during training is applied to the test videos. The model independently predicts quality scores for six sub-videos generated by the sliding window, and the final quality score is calculated as the average of these predictions. 

\subsection{FR-CompressedVQA-HDR - Winning team for FR Track}

\begin{figure*}[!t]
    \centering
    \includegraphics[width=0.75\linewidth]{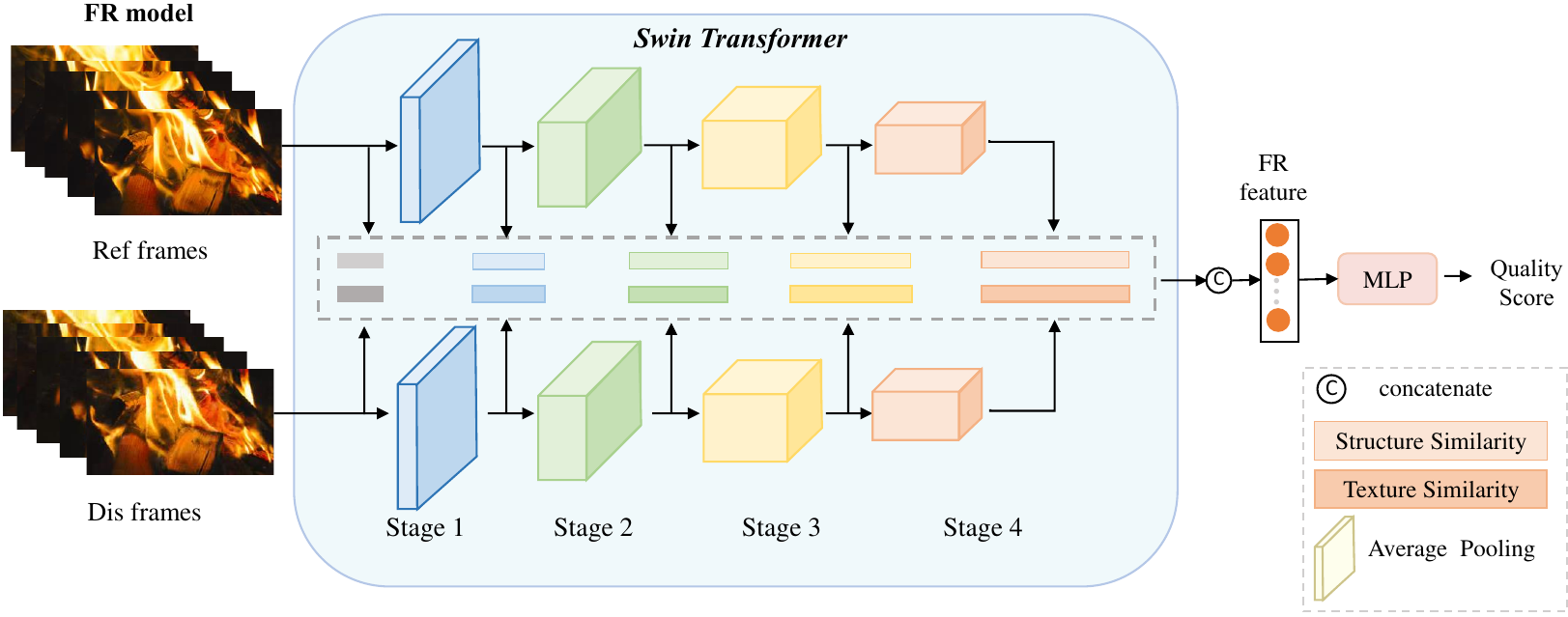}
    \caption{Architecture of the CompressedVQA-HDR model. It extracts Swin Transformer features from reference and distorted frames, computes similarity statistics, and regresses a frame-level score.}
    \label{fig:compressedvqa-hdr}
\end{figure*}

The FR track winning method, FR-CompressedVQA-HDR~\cite{ecnu} proposed by ECNU-SJTU VQA Team, builds upon prior work in SDR UGC-oriented VQA model (CompressedVQA~\cite{sun2021deep}) and extends it to handle HDR content.  This HDR extension aims to generalize quality assessment to HDR content by computing deep textural and structure similarities~\cite{ding2020image} between reference and distorted frames using Transformer-based representations. The model is designed to robustly evaluate visual fidelity under the complex luminance and contrast dynamics of HDR videos, and to generalize well by learning perceptual correspondence in feature space.

As shown in Fig.~\ref{fig:compressedvqa-hdr}, FR-CompressedVQA-HDR comprises two main components: a quality-aware feature extraction module and a quality regression module. For each pair of preprocessed reference and distorted frames $z^r_i$ and $z^d_i$, deep hierarchical features $f^r_i$ and $f^d_i$ are extracted from multiple stages of a Swin Transformer~\cite{liu2021swin} pretrained on ImageNet. At each stage, the model computes \textit{texture similarity} via global mean alignment:
\begin{equation}
f_T(f^r_i, f^d_i) = \frac{2\mu^r_i \mu^d_i + c_1}{(\mu^r_i)^2 + (\mu^d_i)^2 + c_1},
\end{equation}
and \textit{structure similarity} via variance and covariance statistics:
\begin{equation}
f_S(f^r_i, f^d_i) = \frac{2\sigma^{r,d}_i + c_2}{(\sigma^r_i)^2 + (\sigma^d_i)^2 + c_2},
\end{equation}
where $\mu^r_i$, $\mu^d_i$ and $\sigma^r_i$, $\sigma^d_i$ are the global means and standard deviations of the feature maps respectively, $\sigma^{r,d}_i$ denotes the global covariance between them, and $c_1$, $c_2$ are constants for numerical stability. Similarity features from all stages are concatenated into a final quality-aware vector $f_i$, which is passed through a two-layer MLP to predict the frame-level quality score $\hat{q}_i$. The final video-level score $\hat{q}$ is obtained by averaging predictions over all sampled frames.

The model takes as input a pair of reference and distorted HDR videos. To reduce computational cost while preserving perceptual information, frames are sampled at $1$ fps and spatially downsampled using bicubic interpolation to a fixed resolution (\textit{i.e.,} $384\times384$). These preprocessed frames are used for both training and inference. FR-CompressedVQA-HDR is trained using a two-stage strategy to enhance generalization to HDR content. First, the model is pretrained on the Compressed UGC VQA dataset~\cite{ugc2021challenge}, which contains $6,400$ UGC video clips for training and $800$ for validation. Subsequently, the model is fine-tuned on the HDR-VQA dataset released by the ICME 2025 Grand Challenge~\cite{ugc2025challenge}, containing 180 HDR and 180 SDR videos. The dataset is split into training and testing sets with an $8:2$ ratio based on video content. Training is performed using the Adam optimizer with a learning rate of $1 \times 10^{-4}$, a batch size of $6$, and a total of $10$ epochs for pretraining followed by $30$ epochs for fine-tuning. The PLCC loss function is used to optimize the correlation between predicted and ground-truth quality scores.


\subsection{cvvdpMITransformer \& cvvdpMISaliency}

\begin{figure*}[!t]
    \centering
    \includegraphics[width=0.9\linewidth]{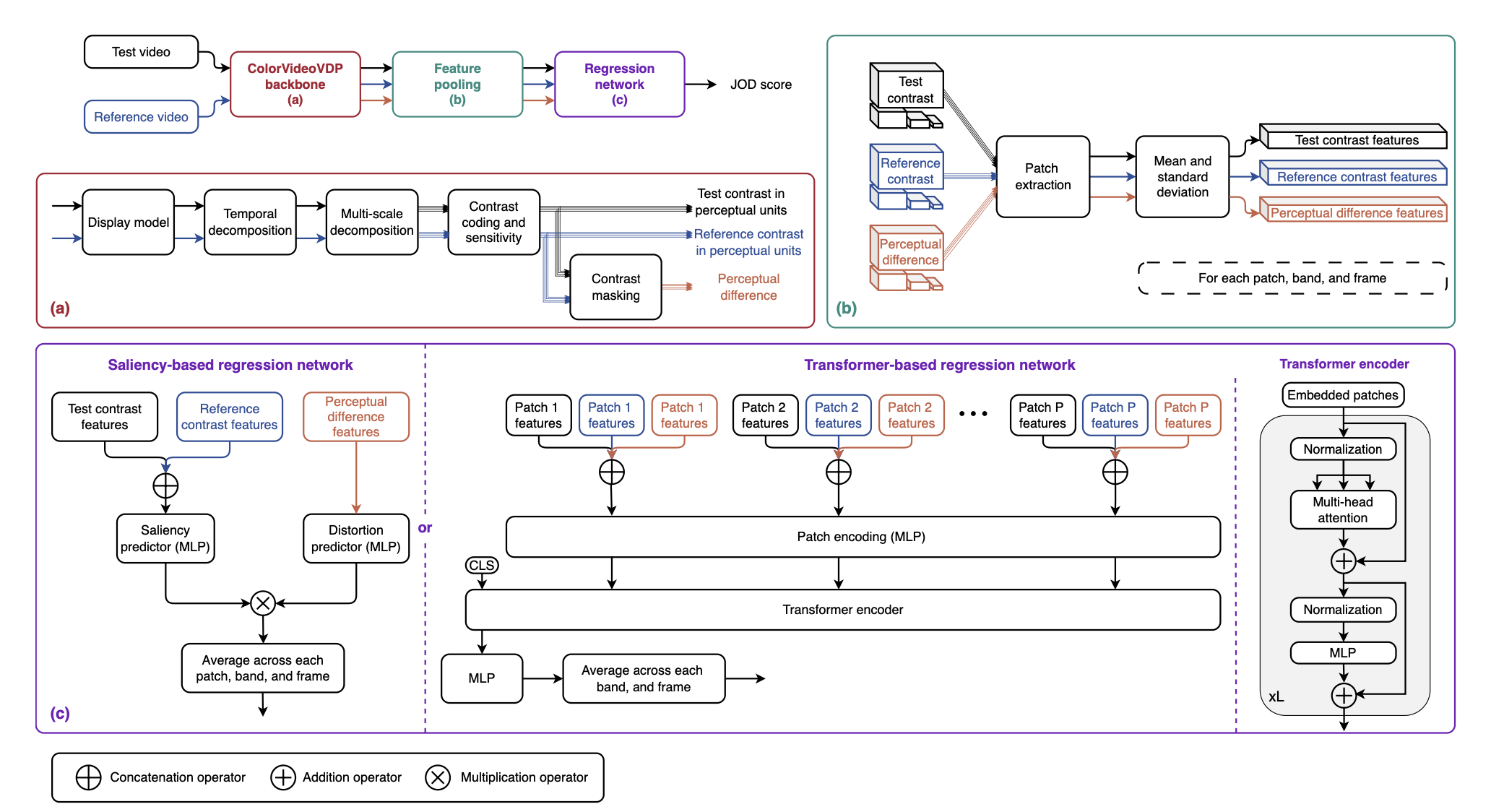}
    \caption{Architecture overview of the proposed ColorVideoVDP-ML. (a) provides details on the ColorVideoVDP metric that is used as a backbone. (b) explains how the features are pooled from the contrast information extracted from ColorVideoVDP. (c) showcases the regression of two architectures.}
    \label{fig:colorvideovdp-ml}
\end{figure*}

ColorVideoVDP-ML~\cite{colorvideo-ml} is a full-reference video quality metric that combines a classical visual difference predictor with a neural regressor shown in Fig.~\ref{fig:colorvideovdp-ml}. The model is built upon ColorVideoVDP~\cite{colorvideovdp}, a perceptual model that simulates low-level human vision, and extends it with machine learning components to account for high-level semantic factors influencing perceived video quality. 

The model first processes both reference and distorted videos through the ColorVideoVDP backbone, which simulates early stages of human visual perception using temporal decomposition, multi-scale filtering, and contrast sensitivity modeling. This produces dense perceptual features such as contrast maps and predicted visibility differences. The perceptual features are pooled spatially into non-overlapping patches corresponding to approximately one visual degree. For each patch, band, and frame, several statistics are computed, resulting in compact but expressive feature vectors. Two different regression heads are explored: Saliency-Based MLP uses two MLPs to predict per-patch distortion magnitude and saliency weight, modulating each patch's contribution based on its importance, while Transformer-Based Regressor learns inter-patch relationships by encoding all patch-level features through a Transformer encoder, followed by a regression MLP head. All scores are aggregated using a weighted average across patches, bands, and frames, yielding a final JOD score.

\subsection{NR-CompressedVQA-HDR}

\begin{figure*}[htbp]
    \centering
    \includegraphics[width=0.8\linewidth]{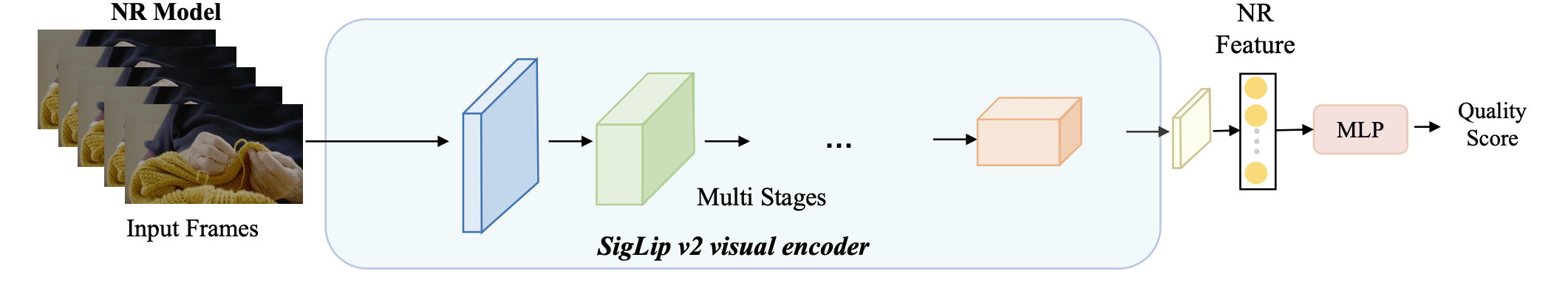}
    \caption{Architecture overview of the NR-CompressedVQA-HDR}
    \label{fig:NR-CompressedVQA-HDR}
\end{figure*}

The NR-CompressedVQA-HDR~\cite{ecnu} proposed by ECNU-SJTU VQA Team adopts a vision-language backbone SigLip v2~\cite{siglip} for no-reference quality prediction of compressed HDR videos. The architecture of NR-CompressedVQA-HDR is shown in Fig.~\ref{fig:NR-CompressedVQA-HDR}. Each distorted frame is processed through the visual encoder of SigLip v2, and the global average pooling of the final-layer feature maps is used as the quality-aware representation. These frame-level features are passed to a lightweight two-layer MLP to predict quality scores, which are averaged across the video to obtain the final prediction.

To address data scarcity in HDR datasets, the model is trained using an iterative mixed-dataset training (IMDT) strategy~\cite{imdt} across multiple VQA datasets. The feature extractor is shared across datasets, while separate regressors are maintained to adapt to different label scales. After this multi-dataset training, the model is fine-tuned on the HDR-VQA dataset. 

\subsection{PhoenixVideo}

\begin{figure*}[!t]
    \centering
    \includegraphics[width=0.8\linewidth]{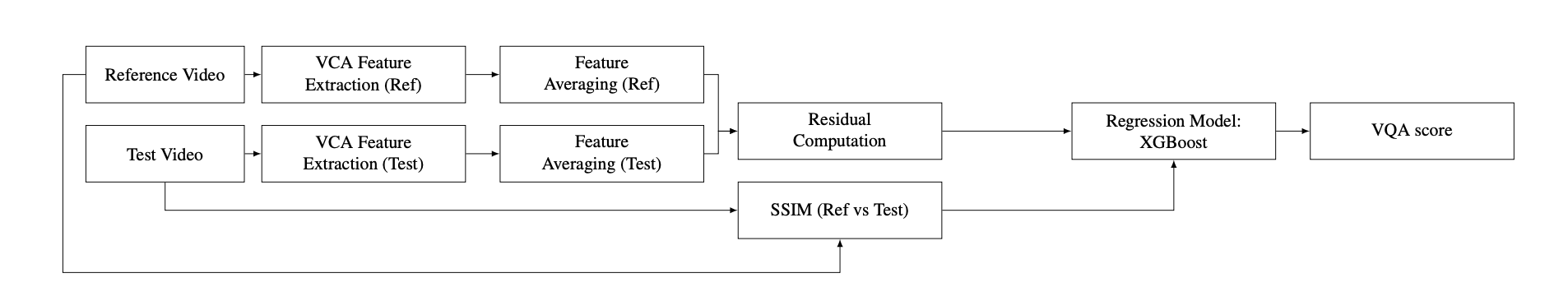}
    \caption{Architecture overview of the PhoenixVideo.}
    \label{fig:phoenixvideo}
\end{figure*}

PhoenixVideo~\cite{phenixVideo} presents a reduced-reference (RR) VQA  method that relies on interpretable, handcrafted features and a lightweight machine learning model. The approach shown in Fig.~\ref{fig:phoenixvideo} begins by extracting seven spatio-temporal features from both the reference and distorted videos using a Video Complexity Analyzer (VCA)~\cite{vca}, which applies 2D-DCT filtering on luma and chroma channels to capture temporal and spatial texture characteristics. These include statistics of luma texture strength, motion energy, chroma activity, and directional content. To model perceptual degradation, residual features are computed by subtracting distorted from reference features. Additionally, the method incorporates intra-frame SSIM~\cite{ssim} mean and standard deviation values, extracted directly from distorted frames, to provide complementary information on local structural integrity. The combined feature set is pooled temporally and passed into an XGBoost regressor trained to predict quality scores. This fully interpretable pipeline is efficient, requires minimal reference information, and avoids the use of deep neural networks, making it well-suited for real-time video quality monitoring and bitrate-adaptive streaming.

\section{Analysis on HDR and SDR scale}
\begin{figure*}[ht]
    \centering
    \begin{subfigure}[b]{0.4\textwidth}
        \includegraphics[width=\textwidth]{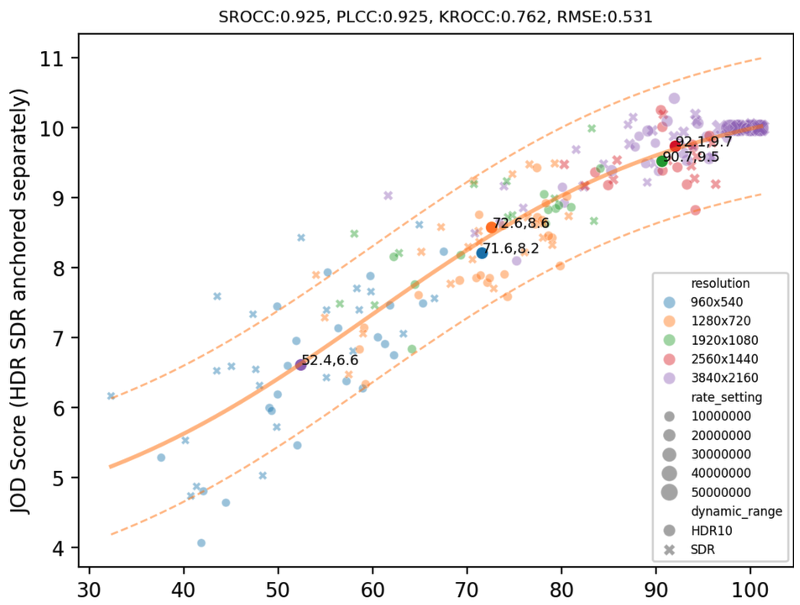}
        \caption{as separate source}
    \end{subfigure}
    \begin{subfigure}[b]{0.4\textwidth}
        \includegraphics[width=\textwidth]{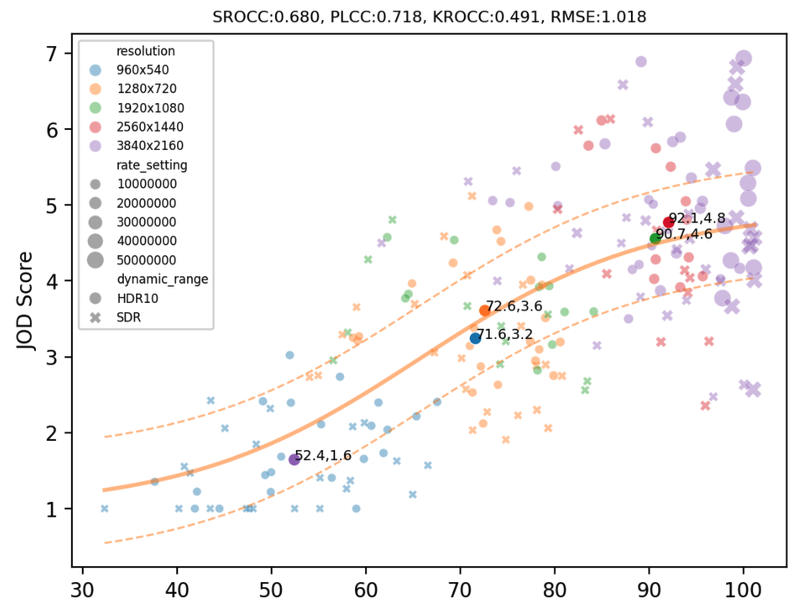}
        \caption{as same source}
    \end{subfigure}
    \caption{EQM\_NR vs JoD where SDR and HDR are recovered differently}
    \label{fig:sdr_hdr}
\end{figure*}

In the case of SDR content is derived from HDR via tone mapping, the HDR serves as the source (SRC), and the SDR is the processed video. SDR and HDR content exhibit similar complexity and should be treated as variants of the same content, with their scores recovered together in a pairwise comparison (PC) subjective study. We observed that most full-reference (FR) metrics do not support comparisons across different pixel formats. For those FR metrics that do support it, or for no-reference (NR) metrics, the correlations drop sharply when SDR and HDR are included in the same rating session and their scores are recovered as if they originate from the same source content because viewers can better discern the HDR characteristics of a video in a PC rating session. For example, as shown in Fig. \ref{fig:sdr_hdr}, the SROCC for EQM\_NR drops from 0.92 to 0.68. The same applies to other metrics; e.g., SLCV changes from 0.94 to 0.96, P1204.3 drops from 0.92 to 0.62, and VMAF drops from 0.90 to 0.55. For simplicity, in the challenge, we used the scores in which SDR and HDR were rated as separate content. Models can be trained to better separate HDR and SDR video and capturing resolution-bitrate-dynamic range crossover~\cite{zhu2025videoqualityassessmentresolution} by using dataset build with PC study protocol with joint rating of HDR and SDR.

\section{Conclusion}

This paper presented a summary of the ICME 2025 Grand Challenge on Generalizable HDR and SDR Video Quality Measurement, which aimed to benchmark VQA methods capable of handling both HDR and SDR content under diverse distortion conditions. The challenge attracted a range of solutions, including deep learning-based architectures, multimodal models, and lightweight interpretable pipelines. We hope that this challenge and its accompanying dataset will continue to support future research on video quality assessment across dynamic range formats, and serve as a useful benchmark for developing more generalizable and perceptually accurate VQA models.

\section{Appendix: Teams and Affiliations}
\textbf{Team: SLCV}
\begin{itemize}
    \item Members: Baojun Li
\end{itemize}

\textbf{Team: PhoenixVideo}
\begin{itemize}
    \item Members: Amritha Premkumar, Prajit T Rajendran, Vignesh V Menon
    \item Affiliations: Department of Computer Science, Rheinland-Pfälzische Technische Universität, Kaiserslautern, Germany;
Université Paris-Saclay, CEA, List, F-91120, Palaiseau, France;
Video Communication and Applications department, Fraunhofer HHI, Berlin, Germany
\end{itemize}

\textbf{Team: ECNU-SJTU VQA Team}
\begin{itemize}
    \item Members: Wei Sun, Linhan Cao, Kang Fu, Dandan Zhu, Jun Jia, Menghan Hu, Xiongkuo Min, Guangtao Zhai
    \item Affiliations: East China Normal University, China; Shanghai Jiao Tong University, China.
\end{itemize}

\textbf{Team: Cambridge VDP Team}
\begin{itemize}
    \item Members: Dounia Hammou, Fei Yin, Rafał K. Mantiuk
    \item Affiliations:Department of Computer Science and Technology, University of Cambridge, UK.
\end{itemize}

\textbf{Team: BVI-VQA Team }
\begin{itemize}
    \item Members: Robert Hamilton, Chengxi Zeng, Chen Feng, Yuxuan Jiang, Tianhao Peng, Fan Zhang, David Bull
    \item Affiliations: Visual Information Laboratory, University of Bristol, United Kingdom.
\end{itemize}

\bibliographystyle{IEEEbib}
\bibliography{icme2025}


\end{document}